\documentstyle[emulateapj]{article}

\begin{document}
\submitted{Accepted by The Astrophysical Journal Letters, 1998 September 8}

\title{Spectral Line Distortions in the Presence of a Close-in Planet}

\author{David Charbonneau\altaffilmark{1}, Saurabh Jha\altaffilmark{2}, and
Robert W. Noyes\altaffilmark{3} }
\affil{Harvard-Smithsonian Center for Astrophysics, 
60 Garden St, Cambridge, MA 02138}

\begin{abstract}
We discuss the interpretation of distortions to stellar spectral lines with
particular attention to line bisectors in the presence of an orbiting
planetary companion. We present a simple model whereby light
\emph{reflected} by the companion can cause temporal variations to the
observed line profiles. These distortions have a characteristic
signature depending on the inclination angle of the system. For the
known close-in extrasolar giant planets, the expected amplitude of the
effect might not be far from current detection capabilities.  This
method could be used to directly detect the presence of the companion,
yielding the orbital inclination and hence the planetary mass.
Futhermore, a detection would measure a combination of the planetary
radius and albedo, from which a minimum radius may be deduced.
\end{abstract}

\keywords{planetary systems --- stars: individual (51 Peg, $\tau$ Boo) --- techniques: spectroscopic}
\altaffiltext{1}{dcharbonneau@cfa.harvard.edu}
\altaffiltext{2}{sjha@cfa.harvard.edu}
\altaffiltext{3}{rnoyes@cfa.harvard.edu}

\section{INTRODUCTION}

The interpretation of low amplitude, periodic radial velocity
variations in solar-type stars as being induced by orbiting low mass
companions was recently put into question.  Gray (1997) and Gray \&
Hatzes (1997) claimed to detect distortions to the line profile
bisector (the locus of midpoints of a stellar absorption line from the
core up to the continuum) in one spectral line 
(Fe {\scriptsize I} $\lambda 6253$)
of the star 51 Pegasi at the now well known 4.23 day radial
velocity period (Mayor \& Queloz 1995; Marcy et al. 1997). They stated
that the planet hypothesis could not account for variations
in the line profiles, and proposed the alternative explanation that
51 Peg was undergoing non-radial pulsations.  
Further observations and analyses (Gray 1998; Hatzes, Cochran \& 
Bakker 1998a,b; Brown et al. 1998a,b) failed to confirm the claimed 
period in the line bisector, and the planet hypothesis has emerged as 
the most reasonable explanation of the radial velocity curve.

Nevertheless, the question of interpretation of variations in the line
bisector remains.  Is it a given that the planet explanation is
immediately excluded should intrinsic variations in the line profiles
of 51 Peg be detected at the claimed orbital period?  
Our purpose in this Letter is twofold; first,
to provide a simple model whereby a close-in orbiting companion
produces periodic distortions to the line profiles, and second, to
explore the observational consequences of this effect, the amplitude
of which might be close to current observational limits, and the specific
signature of which is set by the orbital inclination of the system 
and the radius and albedo of the companion.

\section{REFLECTED LIGHT}

The amount of light reflected by the planets of
our solar system is a tiny fraction of the solar
output. However, for the close-in extrasolar giant planets (CEGPs) discovered to
date, this fraction increases substantially due to the proximity of a large
planet to the star.  The semi-major axes of these systems, $\tau$ Boo, 51 Peg, 
$\upsilon$ And and ${\rho}^1$ Cnc, are 0.046, 0.051, 0.057 and 0.11 AU 
respectively (Mayor \& Queloz 1995; Butler et al. 1997).  
An estimate of the flux ratio can be achieved as follows.   
Let $D$ denote the distance from the star to the planet, 
$R_p$ the radius of the planet, $R_s$ the radius of the star, and 
$\alpha$ the angle between the star and the Earth as seen from the planet.  
We assume that $R_p \ll R_s \ll D$ and that each surface element on the 
planet reflects a fraction $\eta$ of the incident flux uniformly 
back into the local half-sky (Lambert's Law).  If $\eta$ does not vary with position,
then it is simply the Bond albedo.  Integration over the sphere yields the ratio 
of the observed flux of the planet at $\alpha = 0$ to that of the star,
\begin{equation}
\epsilon = \frac{2 \eta}{3} {\left(\frac{R_{p}}{D} \right)}^2.
\label{amplitude}
\end{equation} 
For either 51 Peg b ($D = 0.051$ AU and assuming $R_p \simeq 1.3 R_{\rm{Jup}}$ 
(Guillot et al. 1996)) or $\tau$ Boo b ($D = 0.046$ AU and assuming $R_p \simeq 
1.2 R_{\rm{Jup}}$), the result is $\epsilon \simeq \eta \times 10^{-4}$.  

The observed flux ratio will vary with the angle $\alpha$, which is a combination of
the orbital inclination $i$ and the orbital phase $\phi$. 
Here we define the  orbital phase to be the 
one conventionally given by radial velocity measurements; 
for a circular orbit (the only case we consider, 
as tidal effects circularize the orbits of these extremely close companions), 
$\phi = 0$ is taken to be the time of maximum recessional velocity of the star.  
The angle $\alpha \in [0,\pi]$ 
is then given by $\cos{\alpha} = - \sin{i} \, \sin{2 \pi \phi}$.  
Integration of the intensity over the surface of the sphere 
viewed at an angle $\alpha$ yields the phase dependent observed flux ratio,
\begin{equation}
f(\phi,i) = \epsilon \left(\frac{\sin{\alpha} + (\pi - \alpha) \cos{\alpha}}{\pi}\right). 
\label{flux}
\end{equation}
For a discussion of the albedos and phase functions of solar system objects, 
see Harris (1961).
Note that in the expression for
$f(\phi,i)$, we have ignored the possibility of occultation; eclipses 
would have a much more pronounced effect and are ruled out for 
$\tau$ Boo, 51 Peg and ${\rho}^1$ Cnc (Henry et al. 1997; Baliunas
et al. 1997).

Thus CEGPs at moderate to high inclination and of high reflectivity
will produce a photometric modulation of the system 
which will be accessible to upcoming satellite missions, as 
discussed by Charbonneau (1998).  However, photometry does not make use
of the spectroscopic separation of the planet and star.
The key to the method we propose here is in analyzing the spectrum of the
system.  Assuming the planet reflects an essentially undistorted spectrum
of the star, a copy exists in the observed stellar spectrum with
relative amplitude $f(\phi,i)$ but at a location which is 
Doppler shifted due to the
motion of the planet. 

The ratio of the amplitude of the radial velocity variations of the
planet to that of the star, $K_p/K_s$, is simply the ratio of the
mass of the star to the mass of the planet, $M_s/M_p$.  From the
single line spectroscopic orbit, only the minimum mass of the planet, 
$M_p \sin i$, is determined (assuming the stellar mass is estimated via
other means). Thus, only the combination $K_p/\sin i$ is known. For 51
Peg, $K_s \simeq 56 \; {\rm m \, s^{-1}}$ and $M_p \sin i \simeq 4.4
\times 10^{-4} M_s$, so $K_p/\sin i \simeq 130 \; {\rm km \, s^{-1}}$.  
Thus there can be a huge difference in 
velocity amplitude between the stellar and
reflected light signal. 

For a given stellar spectral line, the reflected line will move back
and forth as the phase changes. At $\phi = 0.25$ 
and $\phi = 0.75$ 
the radial velocities of the star and planet will be the same
and the reflected line will lie on top of the stellar line. At $\phi = 0$
and $\phi = 0.5$, the lines will be maximally separated. A single
spectrum with sufficient signal-to-noise would in principle be able to
show both.  We have obtained many high signal-to-noise, high 
resolution spectra of $\tau$ Boo, and are attempting to isolate the 
Doppler shifted reflected spectrum from the orders of magnitude brighter 
stellar spectrum by making use of the known orbital parameters 
supplied by radial velocity measurements (Butler et al. 1997).  
The results of this complementary investigation will be 
presented in a separate paper.
Here we present an alternate method of detection which capitalizes on
the high resolution observations and analysis methods that have
already been developed for studying line bisectors. The motion of the
reflected line through the stellar line will cause a small 
distortion which might be detectable by careful measurements
of the line bisectors or through other techniques.

\section{MODEL AND RESULTS}

We assume that the planet reflects an undistorted spectrum of the
primary, leading to a composite spectrum consisting of a reflected
component greatly reduced in amplitude and Doppler shifted by the
orbital motion of the companion.  We model this process for an
individual isolated spectral line. Typical line bisectors for
solar type stars show a distinctive \textbf{\textsf{C}} shape caused by
granulation in the photosphere (see e.g., Gray 1992).  
Here, we have described the lines as simple
Gaussian profiles of appropriate width and depth for the star in
question.  While we have calculated more detailed models (multiple Gaussian
components to the line) so as to reproduce the constant bisector
\textbf{\textsf{C}} shape, the effect of this is simply to add a time
invariant term to the bisector shape derived from the symmetric
Gaussian model.

The velocity of the planet relative to the star at phase $\phi$ is simply
\begin{equation}
v_{p}(\phi,i) = - K_s \frac{M_s+M_p}{M_p} \cos{2 \pi \phi}
\label{velocity}
\end{equation}
We have modelled the stellar and planetary line profiles as a function of
velocity by
\begin{eqnarray}
s(v) & = & 1 - a_s \exp {(-v^{2}/{\Delta v_s}^{2})} \\ 
p(v,\phi,i) & = & 1 - a_p \exp {(-(v-v_{p}(\phi,i))^{2}/{\Delta v_{p}}^{2})} 
\end{eqnarray}
where we have explicitly allowed for a different planetary line width 
($\Delta v_p$) and depth ($a_p$) due to the rotation rates of the planet 
and star (see below).

\begin{figure}
\plotone{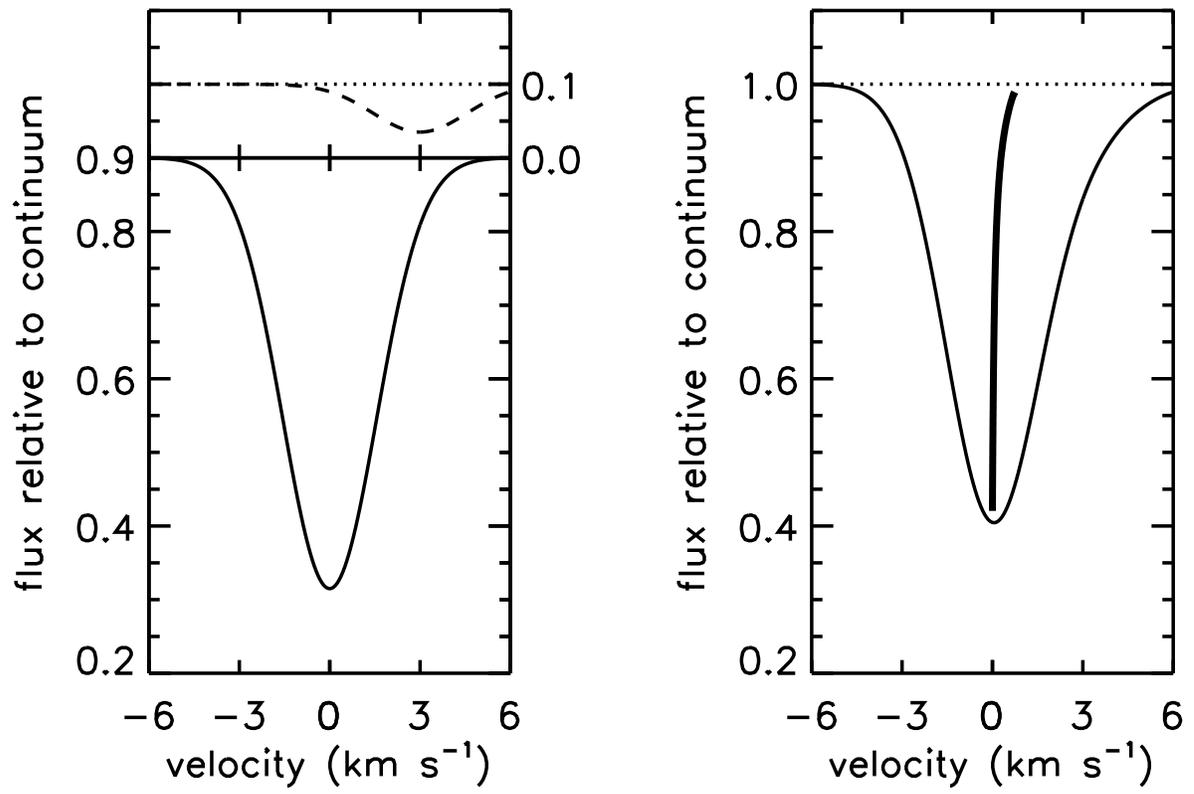}
\caption{The observed line profile is the sum of the stellar 
absorption feature and a time varying, Doppler shifted reflected 
light component.  The 
amplitude of the reflected light has been greatly exaggerated in this 
plot so as to make the resulting distortion to the bisector visible.} 
\label{twolines}
\end{figure}

The flux ratio is given by Equation \ref{flux}, thus the
observed spectrum of the system is 
\begin{equation}
I(v,\phi,i) = \frac{s(v) + f(\phi,i) p(v,\phi,i)}{1 + f(\phi,i)}
\end{equation}
where we have renormalized to the continuum flux level.  The effect of
the reflected light component on the observed bisector is shown
schematically in Figure \ref{twolines}.

For each point in the phase, we deduce the bisector by evaluating the
midpoint between the two halves of the absorption feature at a number
of flux levels, interpolating between sampled points to the
desired flux.  We have also characterized these distortions in terms of two
parameters conventionally used in bisector analyses, the velocity span
and the bisector curvature. The velocity span measures the velocity
difference in the bisector at two flux levels; we have chosen these
levels at two percent below the continuum ($I = 0.98$) and
two percent above the line depth ($I = 1.02 - a_s $). The
bisector curvature is defined to be the velocity span of the top portion 
(between $I = 0.98$ and $I = 1 - ({a_s}/2)$) minus the velocity span 
of the bottom portion (between $I = 1 - ({a_s}/2)$ and 
$I = 1.02 - {a_s}$) of the bisector.

For convenience, we assume $\eta = 1$ in our model calculations.  The
amplitude of the distortion to the bisector (and hence to the velocity
span and curvature) is linear in $\eta$.  The results we present below should
therefore be scaled for a given choice of the reflectivity.

\begin{figure}
\plotone{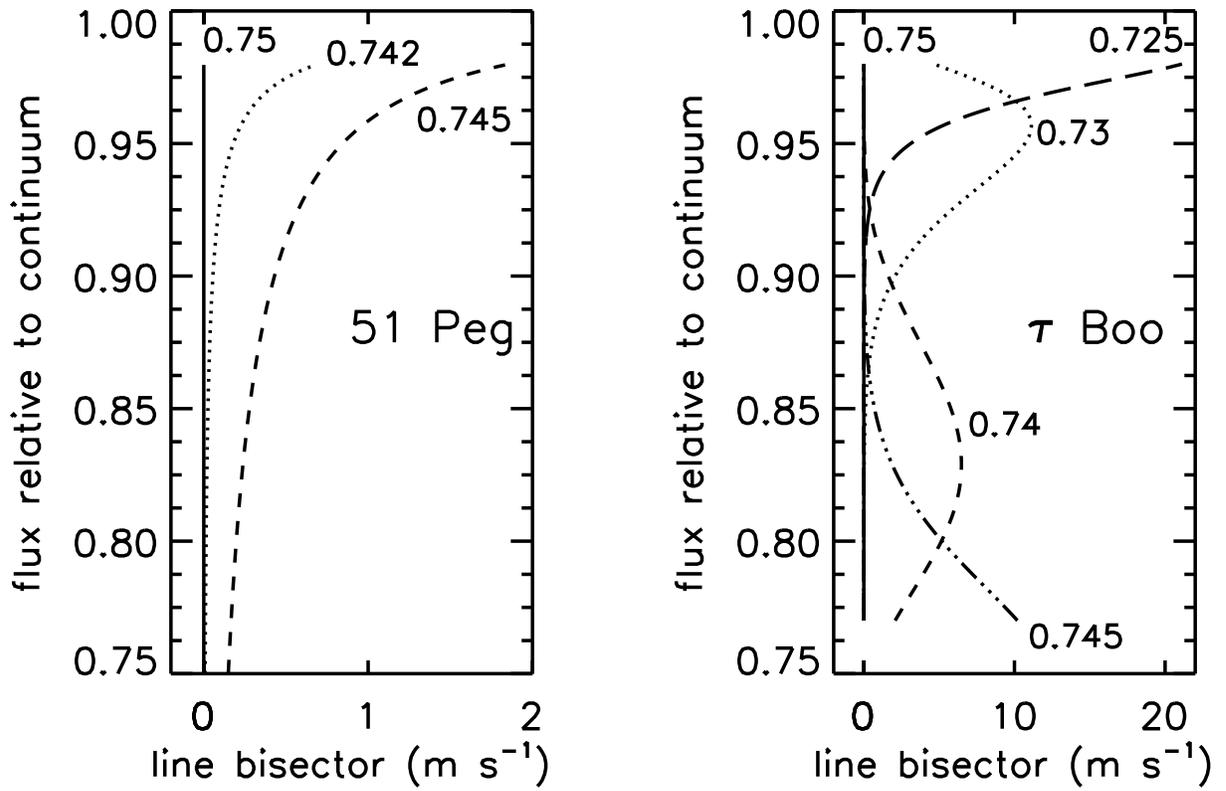}
\caption{Predicted distortions to the bisector of a typical spectral 
line of 51 Peg and $\tau$ Boo for $i = 80 \arcdeg$ and 
$\epsilon = 10^{-4}$.  
The bisectors are labelled by radial velocity phase. 
\label{bisectors}}
\end{figure}

We have computed simulations for 51 Peg with $\epsilon =
10^{-4}$ and a stellar line width of $\Delta v_s = 2.2 \; \rm{km \,
s^{-1}}$.  We assume the reflected light is undistorted, so that
$\Delta v_{p} = \Delta v_{s}$ and $a_{p} = a_{s}$.  We sample the data
simulating a spectral resolution $\lambda/\Delta\lambda = 200,000$
comparable to current high-resolution observations. The predicted
distortions to the bisector for $i = 80 \arcdeg$
are shown in the left panel of Figure \ref{bisectors} for a selection of radial
velocity phases.  The deduced velocity span and bisector curvature are shown 
in the top panels of Figures 3 and 4 respectively.  
The bisector is distorted for the 
fraction of the phase where the stellar and planetary lines overlap and 
the planet is sufficiently illuminated as viewed by the observer.  For a 
near edge-on geometry in the case of 51 Peg, this corresponds to 
roughly 3\% of its 4.23 day period, or about 3 hours, centered on 
$\phi = 0.75$.  Although the stellar and planetary lines also overlap 
at $\phi = 0.25$, there is less distortion as the planet is less 
illuminated as viewed by the observer.  For lower inclinations, the 
fraction of the period in which there is a significant distortion is 
greater, but the amplitude of the distortion is less.

\begin{figure}
\plotone{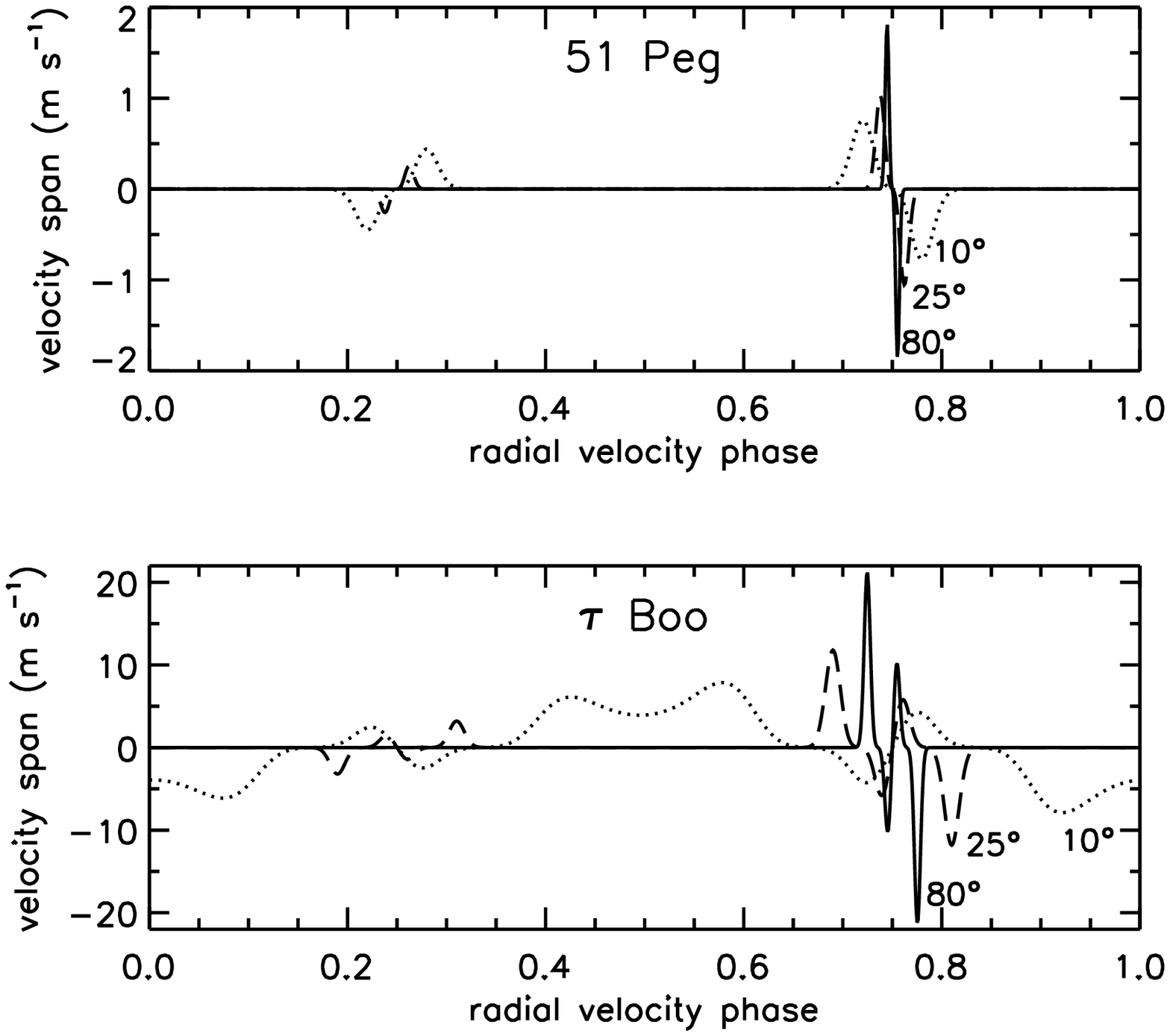}
\caption{The velocity span variations with radial
velocity phase for 51 Peg and $\tau$ Boo, both with $\epsilon = 10^{-4}$.  
The inclinations are $80 \arcdeg$ (solid), $25 \arcdeg$ (dashed) and 
$10 \arcdeg$ (dotted).}
\label{span}
\end{figure}

\begin{figure}
\plotone{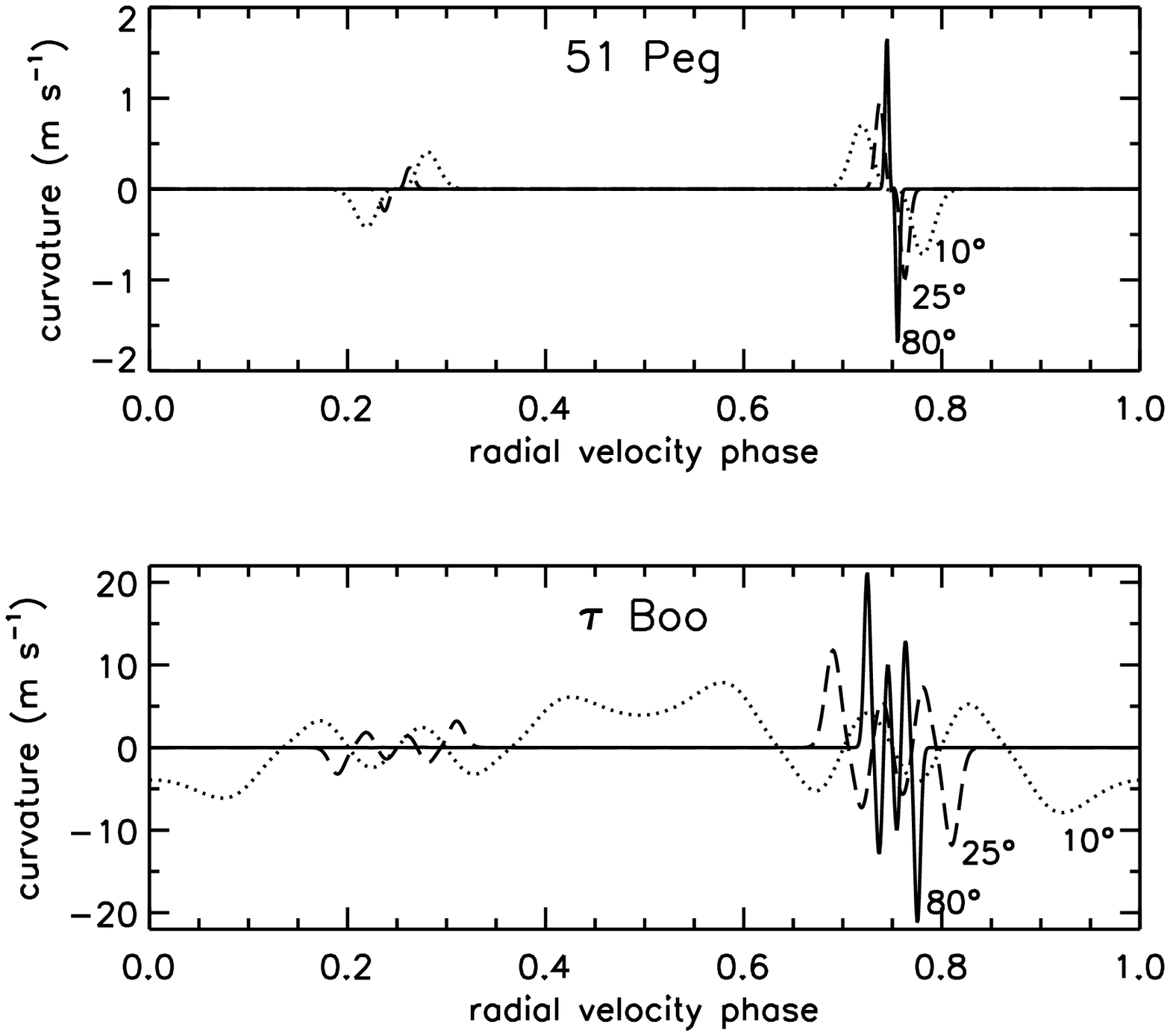}
\caption{The curvature variations with radial velocity
phase for 51 Peg and $\tau$ Boo, both with $\epsilon = 10^{-4}$.  
The inclinations are $80 \arcdeg$ (solid), $25 \arcdeg$ (dashed) 
and $10 \arcdeg$ (dotted).}
\label{curvature}
\end{figure}

We have also modelled this effect for another CEGP 
system, $\tau$ Boo ($K_s = 469 \;
{\rm m \, s^{-1}}$ and $M_p \sin i/M_s = 3.1 \times
10^{-3}$, (Butler et al. 1997)).  Observations of Ca {\scriptsize II} H
$+$ K lines suggest that the rotational period of the star is the same
as the observed orbital period of the companion (Baliunas et
al. 1997), implying that the star and planet are tidally locked.  The 
synchronization timescales for $\tau$ Boo are discussed 
in Marcy et al. (1997).  
If indeed the system is tidally locked, 
the planet does not see the rotational broadening of the spectral
lines of the primary, as there is no relative motion between any point
on the surface of the planet and any point on the surface of the star.
Thus the planet reflects a rotationally unbroadened spectrum of the
primary, with an intrinsic width due to the stellar photospheric
convective motions. An observer then sees this line broadened only by
the planetary rotation, which in this tidally locked scenario is
small.  Thus in the case of $\tau$ Boo, we predict relatively narrow planetary
lines (we adopt $\Delta v_{p} \simeq 4 \; {\rm km \, s^{-1}}$)
superimposed on the much broader stellar lines ($\Delta v_{s} \simeq 15 \;
{\rm km \, s^{-1}}$).  The results of our 
simulated bisector distortions for $i = 80 \arcdeg$ 
are shown in the right panel of Figure \ref{bisectors} 
for a selection of radial
velocity phases.  The deduced velocity span and curvature with phase 
are shown in the bottom panels of 
Figures \ref{span} and \ref{curvature} respectively.  The line is significantly
distorted for the fraction of the phase where the planetary and
stellar lines overlap, which now is substantially increased due to the
high degree of rotational broadening experienced by the lines of the star.  
For an edge-on configuration, this amounts to roughly 10\%, or
about 8 hours of the 3.3 day period, centered on $\phi = 0.75$.
Due to the different line widths, the
deduced bisector varies with a much greater amplitude and in a more
complex manner than in the case of 51 Peg.  The velocity
span and curvature do not well characterize the behavior of the
bisector.  We advocate that the full
bisectors be examined in a search for this effect.

\section{DISCUSSION}

We have shown that the presence of a close-in extrasolar giant planet
can distort the observed spectral line profiles of the system through
the contribution of reflected light from the companion.  For 51 Peg,
assuming $i = 80 \arcdeg$ and $\epsilon = \eta \times 10^{-4}$, 
the amplitude of the distortion is roughly $ 4 \eta \; {\rm m \, s^{-1}} $ and 
$ 3 \eta \; {\rm m \, s^{-1}} $ (peak-to-peak) to the velocity span and 
curvature, respectively.  If $\eta$ is high, the effect 
is commensurate with current observational 
limits:  Hatzes et al. (1998a) combined measurements of five spectral lines
of 51 Peg and achieved a scatter of $ 7 \; {\rm m \, s^{-1}} $
and $ 12 \; {\rm m \, s^{-1}} $ for the velocity span and curvature.
More recent results (Hatzes et al. 1998b) achieve 
a precision of $ 1 \; {\rm m \, s^{-1}} $ for the velocity span and 
$ 4 \; {\rm m \, s^{-1}} $ for the curvature.  In the case
of $\tau$ Boo, assuming that the system is tidally locked 
and that $i = 80 \arcdeg$ and $\epsilon = \eta \times 10^{-4}$, the velocity span and
curvature distortions are both about $ 42 \eta \; {\rm m \, s^{-1}} $
peak-to-peak.  Again, for a high value of $\eta$, 
these are close to the current upper limits 
as measured by Hatzes \& Cochran (1998).  

The amplitude of the velocity span and curvature variations we have 
stated here rely upon measurement of the line bisector to within 2\% of 
the continuum flux level.  Bisector measurements tend to avoid the 
extreme wings of the line profile as $d I / d \lambda$ is small 
and hence the deduced bisector is 
more easily corrupted by noise.  For 51 Peg, the equal linewidths of the 
stellar and reflected spectrum mean that much of the signal in the
velocity span and curvature is a result of the distortion of the bisector 
close to the continuum, as shown in Figure 2.  
For $\tau$ Boo, the different line widths in the tidally locked 
scenario produce a second trough and peak in the velocity span and curvature 
which are insensitive to the height to
which the bisector is measured, because the distortion occurs
away from the continuum.  Note that confirmation of the predicted distortions does 
not require accurate measurement of the \emph{intrinsic} line bisector, only 
precise measurement of the temporal variations to it, and thus it is not 
threatened by contamination from line blends.  Also, isolated lines are not required
as distortions induced by the transit of nearby lines can easily be modelled under the
assumption that the planet reflects a Doppler shifted copy of the stellar spectrum.

The estimate of the reflected light amplitude presented 
in Equation 1 does not consider the underlying source of the albedo, nor 
any angle dependent scattering effects associated with it.  Estimates of 
the albedos of these CEGPs are uncertain as the chemistry of 
these atmospheres is unknown.  The gas giants of the solar system
have Bond albedos ranging from 0.7 to 0.9.  The large reflectivity
is primarily due the presence of ices and clouds high in the atmosphere, and
these could not exist at the much greater effective temperatures of 
CEGPs.  Dominant sources of reflectivity for CEGPs are scattering from dust
and Rayleigh scattering.  Detailed modelling of the 
atmospheric chemistry of these planets indicates that condensates such 
as ${\rm MgSiO_{3}}$ may indeed form at the required height 
(Burrows et al. 1997).  Recent models 
by Seager \& Sasselov (1998), which explicitly solve the radiative transfer 
through a model atmosphere for $\tau$ Boo b, predict an albedo of less than 1\% 
at 5000 \AA.  Although this model is incomplete in that it considers
only one dust species, of interstellar size distribution and solar adundance, it seems
unlikely that albedos as high as those of the solar system gas giants could be reached 
by a dusty CEGP; further investigation is
called for.  An upper limit on the bisector distortions which is sufficiently 
strict to disallow a large value of $\eta$ would provide support for such models.

\section{CONCLUSION}

Via reflected light, a close-in extrasolar giant planet could produce
a distortion to the stellar spectral line bisectors at a level which
is approaching the current observational upper limits, given a sufficiently 
high albedo.  The variation
of the bisector distortions with phase has a very distinctive
signature which depends on the inclination of the orbit of the planet
to the line of sight.  Measurement of such periodic
bisector variations would constitute a direct detection of
the companion, a major advance in the field of extrasolar planet
research.  This would yield the inclination, and hence the mass of the
companion.  It would also measure the product of the albedo and the
square of the radius of the planet, from which a minimum planetary
radius may be deduced.  An estimated albedo would yield a value for the
planetary radius, which can be combined with the measured mass to
calculate such critically important quantities as the surface gravity
and average density.  Conversely, a firm upper limit on such distortions would
place a constraint on the albedos of these 
planets.  Either by detection or by an upper limit, 
an accurate measurement of the line profile bisector would place an important 
observational constraint on models of these close-in extrasolar
giant planets.

\acknowledgements
We wish to thank David Latham, in whose informative seminar this idea
was conceived, and Timothy Brown, Artie Hatzes and Dimitar Sasselov 
for pragmatic comments.  
This work was supported in part by a NSF Graduate Research Fellowship.

\newpage

\end{document}